\documentclass[12pt]{article} 
\usepackage{epsf}
\usepackage{pazha}
\tightenlines
\def\ÐÍ{$\pm$}

\begin{document}

\centerline{\it To be published in ``Astronomy Letters'', 2001, v. 27, n. 6}

\title{\bf LOW STATE OF THE X-RAY BURSTER SLX 1732-304 \\ IN THE 
GLOBULAR CLUSTER TERZAN 1 \\ ACCORDING TO RXTE DATA}  
««
\author{\bf S.V.Molkov, S.A.Grebenev and A.A.Lutovinov}

\affil{
{\it Space Research Institute, Russian Academy of Sciences, Profsoyuznaya
str., 84/32, Moscow, 117810 Russia}\\
}
\vspace{2mm}

\centerline{Received on October 23, 2000}
 
\sloppypar 
\vspace{2mm}
\noindent

Observations of the X-ray burster SLX 1732-304 in the  
globular cluster Terzan 1 with the PCA/RXTE instrument in April 1997 are  
presented. The source was in a low state; its flux in the standard X-ray  
band was half the flux recorded by the ART-P/Granat telescope also  
during its low state. At the same time, its spectrum was softer than the  
ART-P spectrum; it was well described by a power law with a photon index  
of 2.3 without any evidence of a high-energy cutoff.

\clearpage
 
\section*{INTRODUCTION}

\vskip -5pt 

Globular clusters stand out from the Galactic stellar population by a  
high abundance of low-mass binary systems, with one of their components  
being a relativistic, degenerate object --- a neutron star or a white  
dwarf. X-ray sources were discovered in twelve globular clusters (Hut  
et al. 1992). Type I X-ray bursts were observed from most of them,  
which is indicative of the nature of their compact object --- a neutron  
star with a weak magnetic field.

X-ray emission from the region of the globular cluster Terzan 1 was
first detected by the HAKUCHO satellite precisely during bursts
(Makishima et al. 1981; Inoue et al. 1981). The two detected bursts were
identified with a persistent X-ray source later, in 1985, when first
X-ray images of this region were obtained with the XRT telescope of the
SPACELAB-2 laboratory (Skinner et al. 1987). The flux from the
source, designated as SLX 1732-304, was $1.8\times 10^{-10}$
erg~cm$^{-2}$~s$^{-1}$ in the 3-30-keV energy band. Subsequent
observations of several X-ray missions showed that the source's emission
was highly variable in intensity. The most catastrophic decline in flux
was recorded in 1999 by the BeppoSAX observatory. The 2-10-keV flux fell
to the level of $4.8\times 10^{-13}$ erg~cm$^{-2}$~s$^{-1}$ (Guainazzi et al.
1999). The data obtained with the ART-P X-ray telescope onboard the
GRANAT observatory in the fall of 1990 (Pavlinsky et al. 1995, 2001) were
of considerable importance in investigating the variability of SLX
1732-304. The source was detected in two different intensity states ---
a low state similar to that observed by the XRT telescope and a high
state with 3-20-keV fluxes of $1.6~10^{-10}$ and $7.0~10^{-10}$ erg
cm$^{-2}$~s$^{-1}$, respectively.  Variations in intensity were
apparently accompanied by hardness variations: whereas the source during
its high state had the thermal spectrum with a distinct exponential
cutoff at high energies typical of bright low-mass X-ray binaries, its
low-state spectrum could be satisfactorily fitted by a power law with a
photon index $\alpha=1.7$.  Here, we continue to analyze the spectral
states of SLX 1732-304 on the basis of its observations with RXTE.

\section*{OBSERVATIONS}
The PCA instrument onboard the RXTE (Rossi X-ray Timing Explorer)  
orbiting X-ray observatory consists of five identical proportional  
counters with a total area of 6500 cm$^2$, the operating energy range 2-60  
keV, and an energy resolution $<18$\% at 6 keV (Bradt et al. 1993).  
Because of its large area, the instrument is sensitive enough for a  
spectral analysis of emission even from weak X-ray sources to be  
performed. The PCA field of view is restricted by a collimator with a FWHM  
of 1\deg. Depending on peculiarities of the suggested study, observational  
data during their initial onboard reduction can be written in various  
telemetric formats. In the observations discussed here, we use data in  
three formats with a time resolution of 0.125 s, 16 s, and 1 $\mu s$ and  
an energy breakdown into 1, 129, and 256 channels, respectively. By the  
time the paper was written, SLX 1732-304 had been within the PCA field  
of view four times (table 1). The total exposure was eight hours.

\begin{table}[t]

\centering 

{{\bf Table 1.} PCA/RXTE observations of SLX 1732-304 in 1997 \label{tab_obs}}

\vspace{0.2cm}
\hspace{0cm}
\begin{tabular}{ccccc}
\hline
\hline
{Observation}&
{Date}&
{Starting time}&
{Exposure}&
{Mean flux}\\
{}&
{}&
{(UT)}&
{(s)}&
{(mCrab)$ ^a$}\\
\hline \\
{20071-10-01-00}&
{18.04}&
{$06^h02^m08^s$}&
{$\sim 5300$}&
{$2.59 \pm 0.06$}\\
{20071-10-01-01}&
{18.04}&
{$20^h37^m04^s$}&
{$\sim 9200$}&
{$2.70 \pm 0.04$}\\
{20071-10-01-02}&
{19.04}&
{$17^h20^m16^s$}&
{$\sim10800$}&
{$2.75\pm0.04$}\\
{20071-10-01-03}&
{20.04}&
{$01^h04^m48^s$}&
{$\sim3500$}&
{$2.80 \pm 0.09$}\\
\\
\hline
\end{tabular}

\vspace{0.3cm}
$^{a }${\protect\small The 3-20-keV flux}

\end{table}

We see from the table that during two observations, the flux from the
source was approximately the same with an insignificant (10\%) trend
toward its increase. The mean 3-20-keV flux was $6.7\times 10^{-11}$
erg~cm$^{-2}$~s$^{-1}$; i.e., it was almost half the flux recorded by
ART-P in 1990 during the low state of SLX 1732-304. For this estimate,
we subtracted the contribution of the 6.7-keV line of diffuse emission
from hot gas in the Galactic bulge (see below). An analysis of the
source's light curve revealed no appreciable variability of its emission
on time scales of tens and hundreds of seconds. An analysis of the power
spectra obtained from data written with a high time resolution revealed
no variability at frequencies 1-1000 Hz either (the $3\sigma$ limit on
the total power in this frequency range was 15\%).

\section*{EMISSION SPECTRUM}

Since the spectra of SLX 1732-304 measured during different observing
sessions were similar in shape, we performed a detailed analysis for the
mean spectrum. Data written with a 16-s time resolution in the 3-20-keV
energy band were used. The PCA response matrix at lower and higher
energies is known with a large uncertainty. The pulse-height spectrum of
the source is shown in Fig. 1 together with its power-law fit
(histogram). The best-fit parameters are: photon index
$\alpha=2.332\pm0.007$ and intensity at 10 keV
$I_{10}=(2.236\pm0.023)\times10^{-4}$
phot.~cm$^{-2}$~s$^{-1}$~keV$^{-1}$. The interstellar absorption (atomic
hydrogen column density) was fixed at $N_H=1.8\times 10^{22}$ cm$^{-2}$,
which was measured by the ROSAT observatory under the assumption of
solar elemental abundances (Johnston et al. 1995). In the PCA operating
range, the effect of such absorption is essentially weak. As can be seen
from the figure, the measured spectrum exhibits an intense emission line
at energy $\sim6.7$ keV. This line is most likely unrelated to the
source SLX 1732-304 itself, but is a superposition of the diffuse 6.64-,
6.67-, 6.68-, and 6.7-keV lines of Fe XXV, whose ions recombine in
clouds of hot plasma near the Galactic center. Using LAC/GINGA
observations (Yamauchi et al. 1993), we estimated the expected intensity
of the iron-line emission to be $F_{6.7}^{\*}\simeq(6-8)\times10^{-4}$
phot.~cm$^{-2}$~s$^{-1}$ toward the globular cluster Terzan 1.  Direct
measurements of the line intensity in the measured spectrum yielded
$F_{6.7}\simeq(4.92\pm0.22)\times10^{-4}$ phot.~cm$^{-2}$~s$^{-1}$.  The
line has the centroid at energy $6.662 \pm 0.011$ keV and a width of $339
\pm 21$ eV ($\sigma$ in a Gaussian profile). Thus, the observed line can
be entirely attributed to the diffuse emission of ionized iron.

Figure 2 shows that the above slight trend in the source's flux during  
the PCA observations was accompanied by changes in its spectrum. The  
figure presents the ratio of the April 18 (open circles) and 20 (filled  
circles) spectra to the best fit to the mean spectrum. We see that  
maximum changes (up to 40\%) took place at soft energies below 7 keV. We  
also see that there is a small excess of photons at energies $\sim 8$ keV in  
both spectra (or a deficit at energies $\sim10$ keV) compared to the  
power-law fit to the mean spectrum.

\section*{DISCUSSION}

Figure 3 shows the X-ray photon spectrum of SLX 1732-304 reconstructed
from the PCA data. The component attributable to the background diffuse
6.7-keV emission was removed. We neglected by possible contribution of
the diffuse continuum emission. This spectrum was taken during the
source's low state, in which it spends most of its time. Significantly,
the observed spectrum is not just very hard, but it is well described by
a power law over the entire PCA operating range without any evidence of
a high-energy cutoff. The X-ray flux was a factor of $\sim2$ lower than
that recorded by ART-P in 1990, also during the source's low state. At
the same time, the spectrum was steeper: the spectral slope had a photon
index $\alpha\simeq 2.3$, whereas the ART-P spectrum had a slope
$\alpha\simeq 1.7$. Discrepancies in the spectra are clearly seen in
Fig. 3, in which the ART-P spectra of SLX 1732-304 observed in 1990
during its low and high states are indicated by open and filled circles,
respectively. In particular, we see that the ART-P low-state spectrum is
in better agreement with a hard Comptonized spectrum than with just a
power law.  This can serve as evidence that this spectrum originated in
an optically thicker and more opaque plasma than did the PCA spectrum.
It is much more similar to the source's high-state spectrum. In general,
it should be said that the PCA data contribute appreciably to the study
of various spectral states of SLX 1732-304.

\section*{ACKNOWLEDGMENTS}

This study was supported by the Russian Foundation for Basic Research  
(project nos. 99-02-18178 and 00-15-99297).

\pagebreak   

\pagebreak




\begin{figure}[t] 
\hspace{-0.5cm}{
\epsfxsize=170mm
\epsffile[18 344 592 550]{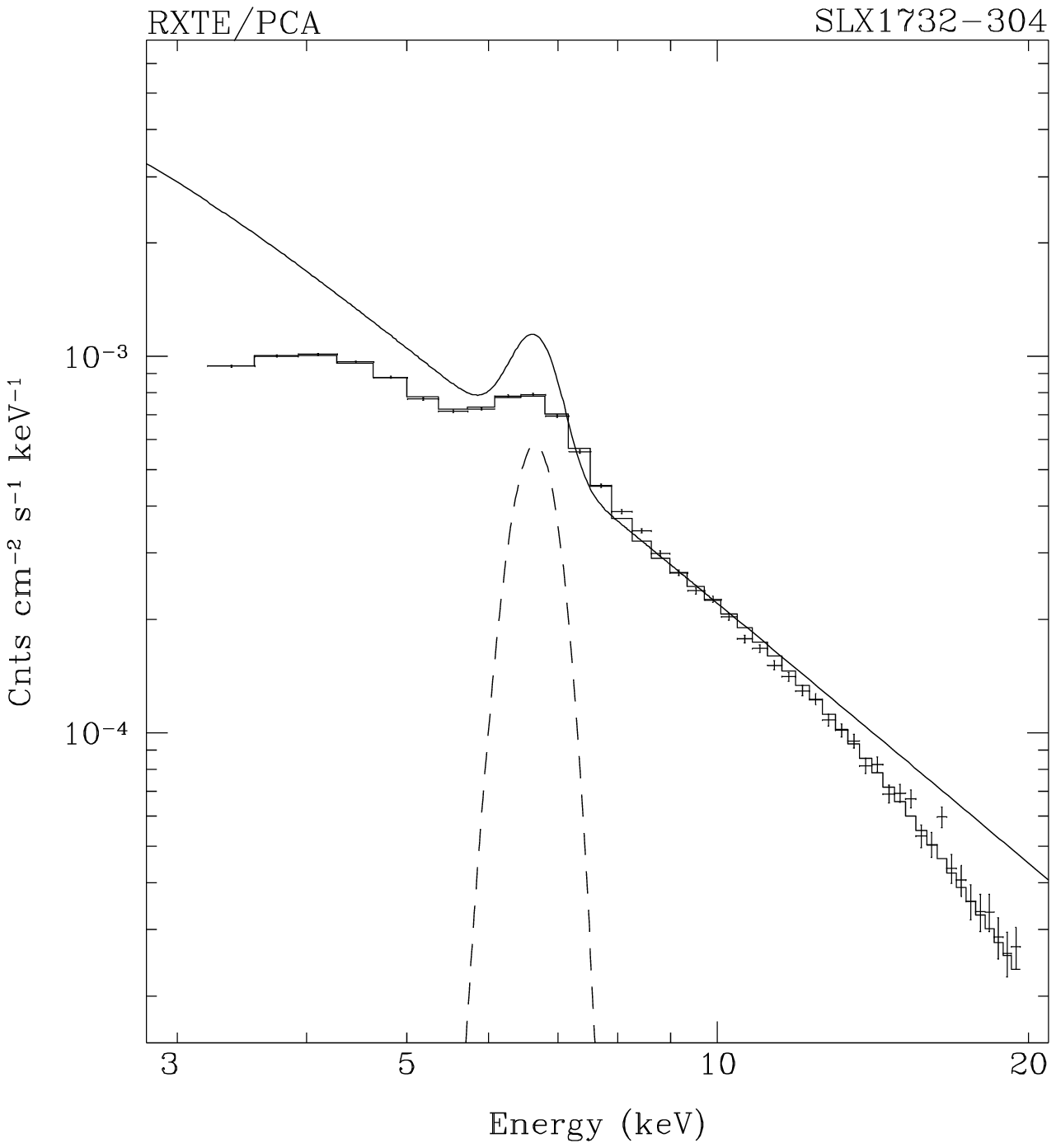}}
\vspace{15mm}

 \caption{\rm PCA pulse-height spectrum of SLX 1732-304 in 1997. The histogram
and the solid line indicate its best fit and the corresponding photon
spectrum, respectively. The 6.7-keV line is attributable to the diffuse
emission of hot plasma in the Galactic-center region.} 
\end{figure}

\clearpage

\begin{figure}[t] 
\hspace{-0.5cm}{
\epsfxsize=170mm
\epsffile[18 344 592 550]{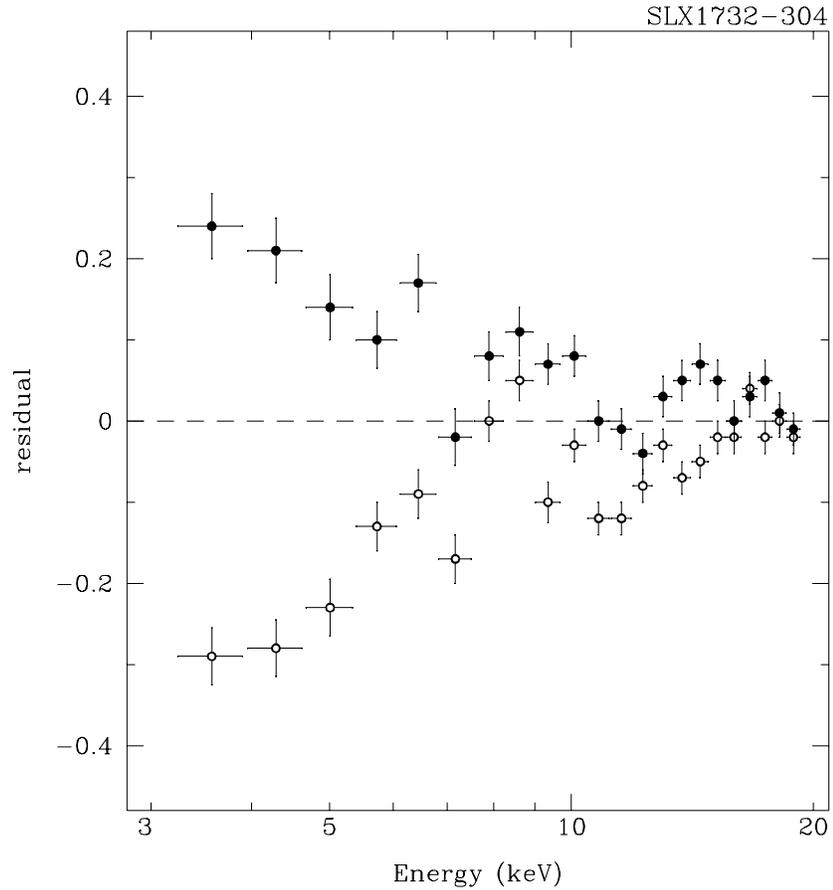}}

\vspace{15mm}

 \caption{\rm Ratio of the pulse-height spectra measured from SLX 1732-304 on
April 18 (open circles) and 20 (filled circles), 1997, to the power-law
fit to the mean spectrum. The figure illustrates the source's spectral
evolution on a time scale of three days of its PCA observations.} 
\end{figure}

\clearpage

\begin{figure}[t] 
\hspace{-0.5cm}{
\epsfxsize=170mm
\epsffile[18 344 592 550]{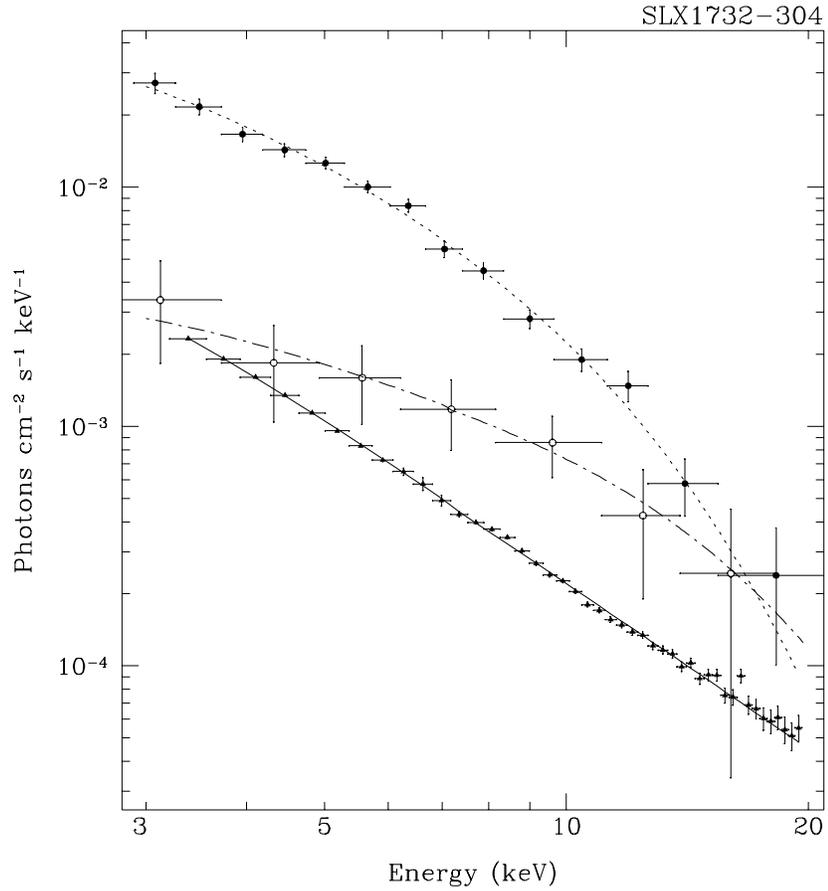}}

\vspace{15mm}

 \caption{\rm Photon spectrum of the persistent X-ray emission (triangles)
observed by PCA from SLX 1732-304 in 1997 during its low state. 
The source's ART-P/Granat spectra (Pavlinsky et al. 2001) measured 
during its low (open circles) and high (filled circles) states in 1990 
are shown for comparison. The results of best approximation of these
spectra with analytic models are shown by curves. } 
\end{figure}

\clearpage

\end{document}